# Determinants of Profitability of Banks: Evidence from Islamic Banks of Bangladesh

- Nusrat Jahan[*]

**Abstract**

This empirical study is conducted on randomly selected six Islamic banks of Bangladesh. This study utilizes widely used Measures of banks' profitability which are Return on Asset (ROA), Return on Equity (ROE) and Return on Deposit (ROD) and these are also commonly suggested tools by Bangladesh Bank to evaluate banks' performance. In addition, this study examined the relationship of ROA with Asset Utilization (AU), Operational Efficiency (OE) and ROD. The result reveals that EXIM Bank Limited is performing very good in terms of all profitability measures ROA, ROE and ROD even though average asset size of Islami Bank Bangladesh Limited is found to be largest among all six Islamic Banks. The result of regression found the explanatory variable ROD is significantly associated with ROA but failed to establish any significant association with operational efficiency and asset utilization.

## 1. Introduction

The dominating industry in the financial sector of Bangladesh is banking industry. Bangladesh Bank (BB) continues to focus on strengthening the financial system of Bangladesh and improving functioning of its various scheduled banks and financial institutions. The importance of bank's financial performance and profitability is appraised both at micro and macro level of economy, since it is one of the important source of finance for enterprises in Bangladesh. The significance of bank's financial performance in the financial sector of Bangladesh led to the undertaking of this research. Since the establishment of Islamic banking system during 1983 in Bangladesh, Islamic banks have gained a footing in the banking industry. Islamic banks not only provide profit-sharing banking facilities, but they also undertake business and trade activities on the basis of fair and legitimate profits. In such banks, ensuring fair practices in dealings with customers and shareholders takes centre stage, more so than in conventional banking where much fair practice needs to be imposed by external regulation (Haron, 1996).

---

[*] The author is Assistant Professor, School of Business, Independent University Bangladesh, Chittagong. Views expressed in this article are the author's own.

The basic aim of any bank is generating profit, which is essential requirement for conducting any business (Bobáková, 2003). To measures profitability of the banks', the common criteria used by BB are Return on Asset (ROA) and Return on Equity (ROE) (Bangladesh Bank, *Annual Report, 2011*). In addition to this, another important profitability measure is Return on Deposit (ROD) (Shrivastava, 1979). The profitability of any bank depends to a great extent on the asset utilization and operational efficiency. The level of bank's resources invested in earning asset contributes to the increased income and profitability of that bank. Besides, the amount of assets that constitute credits and investments is an important asset management decision. Furthermore, income or profit of a bank is the result of revenue function and cost function. The revenue function shows that the total income of a bank deriving from the service rendered by the bank and the cost function shows total expenses incurred in producing any service rendered by the bank (Khan 2008). Hence, efficient management of these functions or operations would lead to increase profitability for the bank. The aforesaid discussion led to the assumption that there is a measurable linkage of profitability of a bank with its asset management and operational efficiency. Henceforth, this study proposes to evaluate the profitability of sample Islamic banks based on common profitability indicators. Furthermore, this paper also investigates the relationship of one of the profitability measure, i.e. ROA with operational efficiency, asset utilization and ROD.

## 2. Banking Industry of Bangladesh

The financial system of Bangladesh is dominated by banks. The banking sector of Bangladesh comprises of Bangladesh Bank as central Bank and four categories of scheduled banks. There are four State-owned Commercial Banks (SCBs), four State-owned Development Financial Institutions (DFIs), thirty Private Commercial Banks (PCBs) and nine Foreign Commercial Banks (FCBs).The total number of banks is 47 as of 2011. These banks had a total number of 7961 branches during 2011. The maximum number of branches is held by SCBs followed by PCBs, DFIs and FCBs. Structure of the banking sector with breakdown for each category is shown in Table-1.

Table 1: Banking System Structure, 2011

(in Billion Tk.)

| Bank Types | Number of Banks | Number of Branches | Total Assets | % of Industry Assets | Deposits | % of Industry Deposits |
|---|---|---|---|---|---|---|
| SCBs | 4 | 3437 | 1629.2 | 27.8 | 1235.6 | 27.4 |
| DFIs | 4 | 1406 | 328.8 | 5.6 | 214.4 | 4.8 |
| PCBs | 30 | 3055 | 3524.2 | 60.0 | 2787.5 | 61.8 |
| FCBs | 9 | 63 | 385.4 | 6.6 | 272.2 | 6.0 |
| Total | 47 | 7961 | 5867.6 | 100.0 | 4509.7 | 100.0 |

**Source:** Bangladesh Bank, Annual Report 2012

In 2011, PCBs held 60 percent of the total industry assets whereas SCBs held 27.8 percent. The least amount of share of industry assets is held by FCBs and DFIs which are 6.6 percent and 5.6 percent respectively. The PCBs share of total industry deposits is 61.8 percent during 2011 which was 27.4 for SCBs, 6 percent for FCBs and only 4.8 percent for DFIs. These data indicates that the performance of PCBs in terms of deposit mobilization and investment in assets is superior compared SCBs, FCBs and DFIs. However, it should be noted that maximum number of scheduled banks fall under the PCBs category, therefore, PCBs holding of maximum share of deposits and assets in the banking industry is justified.

## 2.1 Islamic Banking

Bangladesh entered into Islamic banking system alongside the conventional interest based banking system in 1983. Table 2 reports the number of Islamic banks and Islamic banking branches operating during 2011 and also the market share of the Islamic banking sector in terms of financing, deposits and liquidity of the total banking system. During 2011, out of 47 banks in Bangladesh, 7 PCBs are operating as full-fledged Islamic banks and 16 conventional banks including 3 FCBs are involved in Islamic banking through Islamic banking branches. The full-fledged Islamic banks are Islami Bank Bangladesh Limited, Shahjalal Islami Bank Limited, Al-Arafah Islami Bank Limited, First Security Islami Bank Limited, ICB Islamic Bank Limited, Export Import Bank of Bangladesh Limited and Social Islami Bank Limited.

**Table 2: Comparative Position of the Islamic Banking, 2011**

(in Billion Tk.)

| Particulars | Islamic Banks | Dual Banking | Islamic Banking Sector | All Banking Sector |
|---|---|---|---|---|
| Number of Banks | 7.0 | 16.0 | 23.0 | 47.0 |
| Deposits | 751.2 | 56.2 | 818.9 | 4484.4 |
| Investments (Credit) | 693.0 | 45.8 | 738.8 | 3642.6 |
| Credit Deposit Ratio | 90.9 | 81.4 | 90.2 | 79.7 |
| Liquidity: Excess (+) Shortfall (-) | 31.0 | 0.5 | 31.5 | 358.5 |

**Source:** Bangladesh Bank, Annual Report 2012

Total deposits of the Islamic banks and Islamic banking branches of the conventional banks stood at Tk. 818.9 billion at the end of December, 2011 and this was 18.3 percent of deposits of the total banking system. During 2011 total financing by the Islamic banks and the Islamic banking branches of the conventional banks stood at Tk. 738.8 billion which was 20.3 percent of the credit of the total banking system of the country. Excess liquidity maintained by Islamic banking sector was Tk. 31.5 billion during 2011 and this was 8.8 percent of the excess liquidity maintained by the total banking system. During 2011 credit to deposit ratio of Islamic banking sector stood at 90.2 percent which is higher compared to the ratio of 79.7 percent maintained by the total banking system. Therefore, this data indicates, Islamic banking sector has ample contribution in the total growth of the banking industry of Bangladesh.

### 3. Literature Review

Profitability is an important measure of financial performance for any bank. Analysis of profitability of a bank provides an insight into the effective utilization of assets. Although, there are various measures of earnings and profitability, the best and widely used indicator is ROA, which is supplemented by ROE and ROD.

Loans and investment in securities are a bank's assets and are used to provide most of a bank's income. The ROA is a financial ratio used to measure the relationship of earnings to total assets. It is a ratio of net earnings divided by total assets. According to DuPont Analysis, ROA indicates both income management and cost management of banks by including both asset utilization ratio and net profit margin ratio. Asset utilization ratio is measured by total operating revenue to total assets and net profit margin ratio is measured by net earnings to total operating revenue. Therefore, ROA assesses how efficiently bank is managing its revenues and expenses and also reflects the bank managements' ability to generate profits by using the available financial and real assets (Clark *et al.* 2007 and Lopez 1999). Numerous studies have been found to use ROA as a criterion for measuring profitability (Peters et al. 2004 and Tarawneh, 2006).

To extend credit or loan and to invest in securities, a bank must have money, which comes primarily from the bank's owners in the form of equity capital, from depositors, and from money that it borrows from other banks or by selling debt securities. Credits and investments are important sources of both interest income and non-interest income of the firm. Hence, this study employs ROE as a measure of profitability, which is a ratio of net earnings to shareholders' equity. This ratio indicates to what extent bank is using equity fund to produce earnings. The growth in ROE depends on how actively and efficiently the bank is managing its equity capital. Peters et al. (2004) and Tarawneh (2006) have used the ratio as an indicator of profitability. According to DuPont Analysis, ROE equals ROA multiplied by Equity Multiplier (EM), which indicates the usage of leverage and effect of leverage on ROE of a bank (Clark et al. 2007 and Lopez 1999).

ROD is considered to be better index of profitability in the same way as return on sales for non-banking companies. It is a ratio of net earnings to total deposits (Shrivastava, 1979). This ratio shows percentage return on each dollar of customers' deposit. In another words, it indicates the effectiveness of bank in converting deposits into net earnings (Rosly and Bakar 2003 and Tanaweh, 2006).

Furthermore, profitability is considered as an index of operational efficiency of banks (Shrivastava1979). Efficiency ratio is a popular tool used by bank financial analyst to evaluate the ability of bank to manage its costs. Operational efficiency of a bank is calculated by the ratio of non-interest expense to total operating income. It measures the level of non-interest expense needed to support one dollar of operating income, consisting of both interest income and non-interest fee income (Hays et al., 2009).

Investigating the financial performance and determinants of bank profitability has been one of the popular topics among researchers in banking studies. Till date, researchers have managed to examine and identify various factors that have a significant influence on banks performance and profitability. The literature divides the determinants of bank profitability into two categories, namely internal and external. Internal determinants of profitability, which are within the control of bank management, are basically financial statement variables. External variables are those factors that are considered to be beyond the control of management of a bank. Among the widely discussed external variables are competition, regulation, concentration, market share, ownership, scarcity of capital, money supply, inflation and size (Haron, 2004).However, current study focuses on only internal factors or financial statement variables.

Begrer (1995) examined the relationship between ROE and the capital asset ratio of US banks and found them to be positively related. Positive correlation between ROE and capital has been evident in many past studies (Keeley and Furlong, 1990; Naceur, 2003; and Kwan and Eisenbeis, 2005). Empirical evidence form Naceur and Goained (2001) indicate that the best performing banks are those who have maintained a high level of deposit accounts relative to their assets and this lead to higher return on assets.

Guru et al. (2002) found efficient expenses management as one of the significant factor in explaining Malaysian banks' profitability. Bashir (2003) found Islamic bank's profitability measures respond positively to the increases in capital and negatively to loan ratios. Haron and Azmi (2004) reported in their study that liquidity, deposit items and asset structure, inflation and money supply has statistically significant impact on profitability of Islamic banks. Tarawneh (2006) found that financial performance of the Omani banks was strongly and positively influenced by the operational efficiency and asset management, in addition to the bank size. Naceur and Goaied (2008) examine the impact of bank characteristics, financial structure, and macroeconomic conditions on Tunisian banks' net-interest margin and profitability during the period of 1980 to 2000.The study suggest that banks that hold a relatively high amount of capital and higher overhead expenses tend to exhibit higher net-interest margin and profitability levels, while size is negatively related to bank profitability.

Kosmidou (2008) examined the determinants of performance of Greek commercial banks during the period 1990-2002. This study found that profitability is positively associated with well capitalized banks and lower cost to income ratios. This study also suggests that the growth of Gross Domestic Product (GDP) is positively related to bank profitability, while inflation rate is negatively related to bank profitability during the period under study. More recently, Sufian and Habibullah (2009) reported in their study that bank specific characteristics in particular loan intensity credit risk and cost have positive and significant impacts on profitability of Bangladeshi banks, while non-interest income exhibits negative relationship with bank profitability. This study found that size has a negative impact on Return on Average Equity (ROAE) while it has positive impact on Return on Average Assets (ROAA) and Net Interest Margin (NIM)

The aforesaid discussion indicates that there is abundance of literature on banks' performance studies and on determinants of profitability, but these studies are mostly confined to conventional banks. Up to this date, there has been little research in Bangladesh on the profitability of Islamic banks. The present research is conducted considering the importance of Islamic banks' role in both micro and macro level economy of Bangladesh. Henceforth, this research contributes to the existing literature on banking studies written in the context of Bangladesh.

## 4. Objectives of the Study

The objectives of this study are as follows:

i. To evaluate the profitability position of selected banks with profitability ratios- ROA, ROE and ROD.

ii. To examine ROA and ROE of selected banks through DuPont Analysis.

iii. To examine the relationship of one of the profitability measures, i.e., ROA with operational efficiency, asset utilization and ROD.

## 5. Methodology of the Study

The sample consists of six Islamic banks, which are randomly selected for this study and constitutes about 86% of the seven full-fledged Islamic banks of Bangladesh. These banks are Islami Bank Bangladesh Limited (IBBL), Shahjalal Islamic Bank Limited (SJIBL), Al-ArafahIslami Bank Limited (AAIBL), First Security Islami Bank Limited (FSIBL), ICB Islamic Bank Limited (ICBIBL) and Export Import Bank of Bangladesh Limited (EXIMBL). This study also aims to examine the relationship of ROA with operational efficiency, asset utilization, and ROD. This research primarily focuses on secondary data collected from the annual reports of selected Islamic banks. The present study covers the relevant data for the period 2008-2012 only.

The profitability measures used in this study are ROD, ROA and ROE. In addition, ROA and ROE are also decomposed with the help of DuPont Analysis. Five years average is calculated for all the selected profitability measures for analyzing profitability and examining the statistical association between dependent and independent variables. In examining empirical relationship between ROA and three independent or explanatory variables, named operational efficiency, asset utilization and ROD, correlation and regression have been used Simple correlation coefficient is used to test the linear relationship among all the dependent and independent variables. Furthermore, to determine the significance of a correlation coefficient, a t-test is performed, which hypothesizes that linear relationship between two variables is zero. The result of correlation and t-test is reported through a correlation matrix table. Multiples regression analysis is used to see how far the explanatory variables are related with ROA. To test the significance of beta coefficients t-test is utilised. In addition to this, ANOVA is used to examine the significance of coefficient of determination that is R-square and to report the fitting of regression equation with the help of 'F' value.

## 6. Findings and Analysis

### 6.1. Comparison of the Bank's Profitability Measures- ROA, ROE, and ROD

According to DuPont Analysis, ROE is decomposed into the profitability of assets indicated by ROA and the leverage of the bank measured by EM. By decreasing equity and increasing leverage, i.e. EM, a bank can increase ROE based on any given level of ROA.

Table 3: ROE: DuPont Analysis (ROE= ROA × EM)

| Ratios | SJIBL | FSIBL | EXIMBL | AAIBL | IBBL | ICBIBL |
|---|---|---|---|---|---|---|
| ROE | 21.31% | 11.70% | 20.38% | 20.15% | 14.41% | -26.42% |
| ROA | 1.86% | 1.46% | 2.24% | 1.92% | 1.34% | -7.83% |
| EM | 11.46 | 7.65 | 9.57 | 10.49 | 10.75 | 3.37 |

**Source:** Author's Calculation

As it is visible from the Table 3, ROE is highest for SJIBL, which is recorded as 21.31%. This result can be justified by the increased use of leverage as reported by the higher EM of SJIBL. The lowest ROE is accounted for ICBIBL, which is -26.42%. This poor performance of ICBIBL in terms of ROE is contributed to the losses incurred from bank's assets (ROA) and less usage of leverage as reflected through the EM in Table-3. Ranking of these banks based on ROE reveals SJIBL to be in first position, EXIMBL to be in second position and the last position is held by ICBIBL.

The DuPont Analysis of ROA for sample Islamic banks is presented in Table-4. DuPont analysis presented in this table suggests ROA evaluates both efficiency of bank in utilizing assets and effectiveness in managing income and expenses indicated by Net Profit Margin (NPM).

Table 4: ROA: DuPont Analysis (ROA= NPM × AU)

| Ratios | SJIBL | FSIBL | EXIMBL | AAIBL | IBBL | ICBIBL |
|---|---|---|---|---|---|---|
| ROA | 1.86% | 1.46% | 2.24% | 1.92% | 1.34% | -7.83% |
| NPM | 33.38% | 22.15% | 36.12% | 34.73% | 25.19% | -289.87% |
| AU | 0.0557 | 0.0659 | 0.0590 | 0.0546 | 0.0532 | 0.0911 |

**Source:** Author's Calculation

Highest ROA reported for EXIMBL is 2.13%, which indicates the greater ability of this bank in generating profit by using its available financial and real assets and enhancing bank's efficiency in saving costs and raising income. In Table-4, highest NPM ratio of 36.12% is also reported for EXIMBL which indicates the bank's efficiency in managing costs and income. Negative ROA is reported for ICBIBL which is -26.42%. This negative ROA is a resultant factor of negative NPM ratio of -289.87%. This is highest negative result among all the six Islamic banks as reported in Table-4. Ranking of these banks based on ROA ratio reveals, EXIMBL to be in first position. The second position is held by AAIBL and the last position would be posted to ICBIBL.

Table 5: ROD Ratio of the Sample Islamic Banks

| Ratio | SJIBL | FBSIL | EXIMBL | AAIBL | IBBL | ICBIBL |
|---|---|---|---|---|---|---|
| Return on Deposit | 2.19% | 0.71% | 2.24% | 2.49% | 1.40% | -10.92% |

**Source:** Author's Calculation

Many financial analysts consider ROD as one of the best measures of bank profitability performance. This ratio reflects the bank's management ability to utilize the customers' deposits to generate profits. In Table-5, highest ROD ratio is reported for AAIBL. Higher ROD ratio is reported for both SJIBL and EXIMBL which are also above 2%. Lowest and negative ROD ratio is recorded for ICBIBL, which is -10.92%.

## 6.2. Association of ROA with Asset Utilization, Operational Efficiency and ROD

**Table 6: Sample Islamic Bank's Descriptive Statistics**

| Variable | SJIBL | FSIBL | EXIMBL | AAIBL | IBBL | ICBIBL |
|---|---|---|---|---|---|---|
| Y(ROA) | 1.86% | 1.46% | 2.24% | 1.92% | 1.34% | -7.83% |
| X1(OE) | 30.17% | 48.47% | 32.43% | 30.40% | 36.70% | 103.13%% |
| X2(AU) | 0.0557 | 0.0659 | 0.0590 | 0.0546 | 0.0532 | 0.0911 |
| X3(ROD) | 2.19% | 0.71% | 2.24% | 2.49% | 1.40% | -10.92% |

**Source:** Author's Calculation

Table-6 provides the descriptive statistics of dependent variable ROA and independent variables-OE, AU and ROD for six Islami banks. It is evident from the table that performance of EXIMBL is at peak in terms of profitability measures ROA and ROD. However, FSIBL's performance is soaring in terms of OE and AU even though average asset size of IBBL found to be largest among all six Islami Banks(Asset size of IBBL is Tk. 342,299.31 whereas that of EXIMBL is only Tk. 99,401.31 million).

**Table 7: Correlation Matrix of All Dependent and Independent Variables**

| | | ROA | OE | AU | ROD |
|---|---|---|---|---|---|
| ROA | Pearson Correlation | 1 | | | |
| | Sig. (2-tailed) | | | | |
| | N | 6 | | | |
| OE | Pearson Correlation | -0.980** | 1 | | |
| | Sig. (2-tailed) | 0.001 | | | |
| | N | 6 | 6 | | |
| AU | Pearson Correlation | -0.949** | 0.978** | 1 | |
| | Sig. (2-tailed) | 0.004 | 0.001 | | |
| | N | 6 | 6 | 6 | |
| ROD | Pearson Correlation | 0.997** | -0.992** | -0.966** | 1 |
| | Sig. (2-tailed) | 0.000 | 0.000 | 0.002 | |
| | N | 6 | 6 | 6 | 6 |

**Correlation is significant at the 0.01 level (2-tailed).
**Source:** Author's Calculation

The values reported in Table-6are used to calculate correlation and conduct regression analysis. The correlation results and results of t-test are shown in Table-7. The correlation matrix table indicates that there exist significant positive relationship between ROA and ROD. However, result shows ROA has significant inverse relationship with OE and AU, which is not expected.

Table 8: Multiple Regression Analysis Result

| Particular | Coefficient | Standard Deviation | T-statistic |
|---|---|---|---|
| Constant | -0.037 | 0.013 | -2.905 |
| OE (X1) | 0.064 | 0.026 | 2.490 |
| AU(X2) | 0.196 | 0.252 | 0.777 |
| ROD (X3) | 1.141 | 0.113 | 10.117 |
| Standard Error | 0.00165675 | | |
| $R^2$ | 99.9% | | |
| $R^2$(Adjusted) | 99.8% | | |

Multiple regression analysis was conducted to examine what affect the independent variables have on the dependent variable ROA. The multiple regression result is shown in Table-8 and the critical value of 't' found at 0.10 level of significance is 2.92. This table shows that beta coefficients of operational efficiency, asset utilization and ROD are positive and the association of ROA with ROD is a statistically significant. Hence this suggests that ROD is significant explanatory variable on increasing ROA.

Table 9: Analysis of Variance

| Source | Degree of Freedom | Sum of Squares | Mean of Squares | F-Statistics |
|---|---|---|---|---|
| **Regression** | 3 | 0.008 | 0.003 | 931.877 |
| **Error** | 2 | 0.000 | 0.000 | |
| **Total** | 5 | 0.008 | | |

To examine whether the regression model as a whole is significant, coefficient of determination, r-square is calculated. Table-8 reports 99.8 percent of the variation in ROA is explained by the regression line. The calculated value of F-ratio 931.877 as reported in Table-9 is higher than the critical value of F-ratio 19.2 at 0.05 level of significance which suggest that the regression model as a whole to be statistically significant.

## 7. Conclusion

This study utilizes widely used measures of banks' profitability, which are ROA, ROE, and ROD and these are also commonly suggested tools by Bangladesh Bank to evaluate banks' performance. In addition, this study evaluates the OE ratio, AU ratio and ROD as a measure of Islamic banks' profitability measured by ROA. The result reveals that performance of EXIMBL is excellent, in terms of all profitability measures ROA, ROE and ROD average asset size data is not reported in this report. The correlation result shows that ROA has significant linear relationship with AU, OE and ROD. The result of regression found the explanatory variable ROD is significantly related with ROA but failed to establish any significant association with OE and AU. The sample size for this study is too small, constituting only six Islamic banks and this is considered as the major shortcoming of present study since many commercial banks are operating Islamic Banking branches. However, the total number of private commercial banks operating as full-fledged Islamic banks in Bangladesh is only seven. Furthermore, the data obtained from the sample banks were only for the period 2008-2012 and five years average for all measures of profitability are calculated, which are also considered as major drawbacks of this study. Therefore, this prompts further researches in future covering a longer time period and using panel data model to investigate these stated measures. Further, future researches can be extended by covering both internal and external measures of bank's profitability. The importance of this study may be viewed as the addition with the existing body of literature on Islamic banking studies. Furthermore, it can also serve as a starting point based on which future related studies can be done in the context of Bangladesh.